\documentclass[twoside]{dis04}
\usepackage{amssymb,citesort,epsfig}
\def\runauthor{J\"{a}ger, Kretzer, Stratmann, Vogelsang}
\def\shorttitle{Gluon Polarization}

\begin{document}

\title{Longitudinal Gluon Polarization in RHIC
Double-Spin Asymmetries}

\author{Barbara J\"{a}ger and Marco Stratmann}
\address{Institut f{\"u}r Theoretische Physik, 
Universit{\"a}t Regensburg, D-93040 Regensburg, Germany}
\author{\underline{Stefan Kretzer} and Werner Vogelsang}
\address{Physics Department, Brookhaven National Laboratory,
and 
RIKEN-BNL Research Center, Upton, New York 11973, USA}
\maketitle

\abstracts{
The longitudinally polarized gluon density 
is probed sensitively in hard collisions of polarized protons 
under the condition that the dominant dynamics are 
perturbative and of leading twist origin. 
First data have recently been presented by 
{\protect{{\sc{Phenix}}}} 
on the double-spin asymmetry $A_{\mathrm{LL}}^{\pi}$
for $\pi^0$ production at moderate transverse momentum 
$p_\perp \simeq 1 \div 4\, {\rm GeV}$ and central rapidity. By means of 
a systematic investigation of the relevant degrees of freedom we show 
that the perturbative QCD framework at leading power in $p_{\perp}$ 
produces an asymmetry that is basically positive definite 
in this kinematic range, 
i.e.~$A_{\rm{LL}}^{\pi}\gtrsim {\cal{O}}(-10^{-3})$.}

\section{Introduction}
The determination of the nucleon's polarized gluon density
is a major goal of current experiments with longitudinally
polarized protons at RHIC~\cite{rhicrev}.
It can be accessed through measurement
of the spin asymmetries
\begin{equation}\label{eq:asydef}
A_{\rm{LL}}=\frac{d\Delta \sigma}{d\sigma}=
\frac{d\sigma^{++} - d\sigma^{+-}}{d\sigma^{++} + d\sigma^{+-}}
\end{equation}
for high transverse momentum ($p_{\perp}$) reactions. 
In Eq.~(\ref{eq:asydef}), $\sigma^{++}$ ($\sigma^{+-}$) denotes
the cross section for scattering of two protons with same (opposite)
helicities. Such reactions can be treated in pQCD
via factorization into parton densities, hard scattering cross sections and,
eventually, fragmentation functions.
Hadronic reactions have
the advantage over DIS that the
partonic Born cross sections involve gluons in the initial
state . They may therefore serve to examine the gluon content
of the colliding
longitudinally polarized protons.
We will here consider \cite{prl} the spin asymmetry 
$A_{\rm{LL}}^{\pi}$ for high-$p_{\perp}$ $\pi^0$ production, 
for which very recently the {\sc{Phenix}} collaboration has presented 
first data~\cite{phenix} at a c.m.s.~energy 
$\sqrt{S}=200$~GeV and central rapidity. 

\section{Hard-Scattering calculation}
\begin{figure}[t]
\begin{center}
\vspace*{-0.6cm}
\epsfig{figure=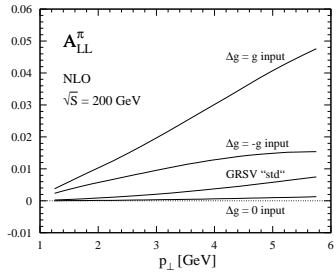,width=0.40\textwidth}
\end{center}
\vspace*{-0.4cm}
\caption{NLO predictions for {\protect $A_{\rm{LL}}^{\pi}$}.\label{fig:all}}
\vspace*{-0.5cm}
\end{figure}

We may write the polarized high-$p_\perp$ pion
cross section as 
\begin{eqnarray} \label{eq:crosec}
\frac{d\Delta \sigma^{\pi}}{d p_\perp d \eta} &=&\sum_{a,b,c}\, 
\int dx_a \int dx_b \int dz_c \,\,
\Delta a (x_a,\mu) \,\Delta b (x_b,\mu) \nonumber \\ 
&\times& \frac{d\Delta \hat{\sigma}_{ab}^{c}
(p_\perp, \eta, x_a, x_b, z_c, \mu)}
{d p_\perp d \eta}\, D_c^{\pi}(z_c,\mu) \,  \; ,
\end{eqnarray}
where $\eta$ is the pion's pseudorapidity. The $\Delta a,\Delta b \;(
a,b=q,\bar{q},g)$ are the polarized parton densities; for instance, 
\begin{equation}
\label{eq:pdf}
\Delta g(x,\mu) \equiv g_+(x,\mu) -
                       g_-(x,\mu) \; ,
\end{equation}
(the sign referring to the gluon helicity in a proton of positive
helicity) is the polarized gluon distribution. 
These parton cross sections start at ${\cal{O}}(\alpha_s^2)$ 
in the strong coupling with the QCD tree-level scatterings:
(i)~$g g \rightarrow gg$, (ii)~$g g \rightarrow q {\bar q}$, 
(iii)~$gq (\bar{q}) \rightarrow g q (\bar{q})$,
(iv)~$q {\bar q} \rightarrow q {\bar q} $, $q {\bar q} 
\rightarrow g g $, $qq\to qq$, $qq'\to qq'$,
$q\bar{q}\rightarrow q'\bar{q}'$. 
The transition of parton $c$ into the observed $\pi^0$
is described by the (spin-independent) fragmentation function
$D_c^{\pi}$. 
All next-to-leading order [NLO, ${\cal O}(\alpha_s^3)$] QCD
contributions to polarized parton scattering are known \cite{JSSV}. 
Corrections to Eq.~(\ref{eq:crosec}) itself are
down by inverse powers of $p_\perp$ and are thus expected 
to become relevant if $p_\perp$ is not much bigger than typical
hadronic mass scales.  
Fig.~\ref{fig:all} 
shows NLO predictions for $A_{\rm{LL}}^{\pi}$,
for various gluon polarizations $\Delta g$ \cite{grsv}. 
Despite the fact that the
$\Delta g$'s used in Fig.~\ref{fig:all} are all very different from 
one another, none of the resulting $A_{\rm{LL}}^{\pi}$ is
negative in the $p_{\perp}$ region we display. We find it interesting
to investigate the question whether this observation is accidental or
systematic and what one could learn from a possible violation of the
apparent positivity of $A_{\rm{LL}}^{\pi}$.
%
\section{Basic observations} 
We first focus on the partonic cross sections. 
Among the reactions (i)-(iii) listed above that have gluons in the initial 
state, process (ii) has a negative {\em partonic} spin asymmetry
${\hat a}_{\rm{LL}}\equiv -1$, while (i) and (iii) both have 
${\hat a}_{\rm{LL}}>0$~\cite{rhicrev}. 
A first guess is, then, to attribute a negative
$A_{\rm{LL}}^{\pi}$ to the negative $gg\to q\bar{q}$ cross
section. However, this expectation is refuted by the numerical 
hierarchy in the partonic cross sections: at $\hat{\eta}=0$ in the 
partonic c.m.s., which is most relevant for 
the {\sc{Phenix}} data, channel (i) is (in absolute magnitude) 
larger than (ii) by a factor of about 160. 
We therefore exclude that the $gg\to q\bar{q}$ channel
is instrumental in making $A_{\rm{LL}}^{\pi}$ negative, and 
we thus have to investigate possibilities within $\Delta g$ itself, 
and its involvement in $gg\to gg$ and $qg\to qg$ scattering. 

\section{A lower bound on $\mathbf{A_{\rm{LL}}^{\pi}}$}
\begin{figure}[t]
\begin{center}
\vspace*{-0.6cm}
\epsfig{figure=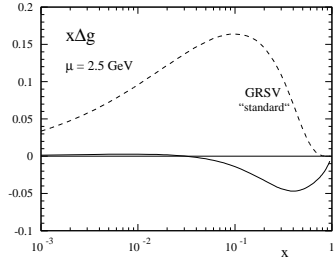,width=0.40\textwidth}
\end{center}
\vspace*{-0.4cm}
\caption{$\Delta g(x,\mu=2.5\,\rm{GeV})$ resulting 
from Eq.~(\ref{dgmin}) (solid) compared to 
GRSV LO ``standard'' 
$\Delta g$
{\protect \cite{grsv}}. \label{fig:dg}}
\end{figure}

We consider the LO cross section integrated
over all rapidities $\eta$. 
It is then convenient to take Mel\-lin moments in $x_T^2$ of the cross section, 
\begin{equation} \label{doublemom}
\Delta\sigma^{\pi} (N) \equiv
\int_0^1 dx_T^2 \left( x_T^2 \right)^{N-1} 
\frac{p_{\perp}^3 d\Delta\sigma^{\pi}}{d p_{\perp}} \; .
\end{equation}
One obtains  (we suppress the scale $\mu$ from now on):
\begin{equation} \label{crosec2}
\Delta\sigma^{\pi} (N) = \sum_{a,b,c} \,\Delta a^{N+1}\,
\Delta b^{N+1}\,\Delta
\hat{\sigma}_{ab}^{c,N}\,D_c^{\pi,2N+3}\; ,
\end{equation}
where the $\Delta\hat{\sigma}_{ab}^{c,N}$ are the $\hat{x}_T^2$-moments 
of the partonic cross sections and, as usual,
$ f^N\equiv \int_0^1 dx\, x^{N-1} f(x)$
for the parton distribution and fragmentation functions. 
We now rewrite Eq.~(\ref{crosec2}) in a 
form that makes the dependence on the moments $\Delta g^N$ explicit:
\begin{equation} \label{quad1}
\Delta\sigma^{\pi} (N) =
\left(\Delta g^{N+1} \right)^2 {\cal A}^N + 
2 \Delta g^{N+1} {\cal B}^N + {\cal C}^N \; .
\end{equation}
Here, ${\cal A}^N$ represents the contributions from $gg\to gg$ and
$gg\to q\bar{q}$, ${\cal B}^N$ the ones from $qg\to qg$, and
${\cal C}^N$ those from the (anti)quark scatterings (iv) above; in each 
case, the appropriate combinations of $\Delta q$, $\Delta \bar{q}$ 
distributions and fragmentation functions are included.  
Being a quadratic form in $\Delta g^{N+1}$, $\Delta\sigma^{\pi} (N)$
possesses an extremum, given by the condition 
\begin{equation} \label{dgmin}
{\cal A}^N \Delta g^{N+1}  = -{\cal B}^N \; .
\end{equation}
The same equation may also be derived
by finding the stationarity condition along a variational approach.
In the following we neglect the contribution from the 
$gg\to q\bar{q}$ channel which, as we discussed 
above, is much smaller than that from $gg\to gg$
for the $p_{\perp}$ we are interested in. The
coefficient ${\cal A}^N$ is then positive, and
Eq. (\ref{dgmin}) describes a minimum of $\Delta\sigma^{\pi} (N)$,
with value
\begin{equation} \label{crsecmin}
\Delta\sigma^{\pi} (N) \Big|_{\rm{min}} =
 -\left({\cal B}^N \right)^2/{\cal A}^N + {\cal C}^N \; .
\end{equation} 
It is straightforward to perform a numerical Mellin
inversion of this minimal cross section:
\begin{equation} \label{inverse}
\frac{p_{\perp}^3 d\Delta {\sigma}^{\pi}}{\d p_{\perp}}
\Bigg|_{\rm{min}} = \frac{1}{2\pi i}
\int_{\Gamma} dN \left(x_T^2\right)^{-N} \, \Delta\sigma^{\pi} (N) 
\Big|_{\rm{min}}\,, 
\end{equation}
where $\Gamma$ denotes a suitable contour in complex-$N$ space.
For the numerical evaluation we use the LO $\Delta q$, 
$\Delta \bar{q}$ of GRSV~\cite{grsv}, the $D_c^{\pi}$ of~\cite{kkp},
and a fixed scale $\mu=2.5$~GeV. We find that the minimal 
asymmetry resulting from this exercise is negative indeed, 
but very small: in the range $p_\perp \sim 1 \div 4$~GeV
its absolute value does not exceed $10^{-3}$. The $\Delta g$
in Eq.~(\ref{dgmin}) that minimizes the asymmetry
is shown in Fig.~\ref{fig:dg}, compared to $\Delta g$
of the GRSV LO ``standard'' set \cite{grsv}. One can see that 
it has a node and is generally much smaller than the GRSV one, except 
at large $x$. The node makes it possible to probe the two gluon densities 
in the $gg$ term at values of $x_a$, 
$x_b$ where they have different sign, so that 
$A_{\rm{LL}}^{\pi}<0$ becomes just barely possible. 

\section{Conclusions}
In its details, the bound in Eq.~(\ref{inverse})
is subject to a number of corrections, however, 
a global analysis \cite{prl}, taking
into account the results from polarized DIS as well confirms 
the somewhat idealized case, as summarized by
Eqs.~(\ref{crsecmin}) and (\ref{inverse}),
that $A_{\rm{LL}}^{\pi}$ is 
basically positive definite 
 $A_{\rm{LL}}^{\pi}\gtrsim {\cal{O}}(-10^{-3})$
in leading twist pQCD
and for $p_\perp$ not largely exceeding $\sim$ 4 GeV.
At the same time,  $A_{\rm{LL}}^{\pi}$ is trivially bounded
from above by saturating positivity through $\Delta g (x) = g(x)$, see the
corresponding curve in Fig.~\ref{fig:all}.
A significantly negative  $A_{\rm{LL}}^{\pi}$ would be indicative
of power-suppressed contributions or non-perturbative effects. 
One possibility might be the population of low $p_\perp$
bins with statistical pions that follow a quasi-thermal exponential
distribution. However, such random pions would have to realize the $J=1$ 
configuration in Eq.~(\ref{eq:asydef}) either through angular momentum
of (Goldstone) pions or the spin of co-produced massive baryons, 
leading one to expect
a positive ${A_{\rm{LL}}^{\pi}}_{\rm nonpert.}>0$.
A quantitative estimate of
this effect will be given elsewhere.

\section*{Acknowledgements}
This work was supported in part by BMBF and DFG, Berlin/Bonn
and by  RIKEN, BNL, 
and the U.S.\ Department of Energy (contract DE-AC02-98CH10886).


\begin{thebibliography}{99}
\bibitem{rhicrev} See, e.g.: 
G.~Bunce {\em et al.},
Annu.~Rev.~Nucl.~Part.~Sci.~{\bf 50}, 525 (2000).
%
\bibitem{prl}
B.~J\"{a}ger {\em et al.},
Phys.~Rev.~Lett.~{\bf 92}, 121803 (2004).
%
\bibitem{phenix} PHENIX Collab.~(S.S.~Adler {\em et al.}), 
{\tt hep-ex/0404027}; and F.~Bauer's contribution to these proceedings.
%
\bibitem{JSSV} D.~de Florian, Phys. Rev. {\bf D67}, 054004 (2003);
B.~J\"{a}ger {\em et al.},  Phys. Rev. {\bf D67}, 054005 (2003).
%
\bibitem{grsv} M.~Gl\"{u}ck {\em et al.}, Phys. Rev. {\bf D63}, 
094005 (2001). 
%
\bibitem{kkp} B.A.~Kniehl {\em et al.}, Nucl. Phys. {\bf B582}, 514 (2000).
%
\end{thebibliography}
\end{document}